# Homegrown Governments:
# Visualizing Regional Governance in the United States

Abdulelah Abuabat
School of Computing and Information
University of Pittsburgh
aaa185@pitt.edu

Steven Johnston
School of Computing and Information
University of Pittsburgh
sdj25@pitt.edu

Mohammed Aldosari
School of Computing and Information
University of Pittsburgh
maa303@pitt.edu

Taylor Neal
School of Computing and Information
University of Pittsburgh
tmn23@pitt.edu

## ABSTRACT

Regional Intergovernmental Organizations (RIGOs) are constituted by the local governments within their respective regions and are supported by the active engagement of the region's community and citizens. Metropolitan Statistical Areas (MSAs), on the other hand, are classified by the federal government based on commuting and commerce patterns. They do not adhere to any local government. The Graduate School of Policy and International Affairs Center for Metropolitan Studies (GSPIA) at the University of Pittsburgh have been researching the boundaries of RIGOs and the characteristics defining them. In this paper, we propose, design, and implement an approach to enhance the current visualization by visualizing two categorical data: RIGOs and MSAs and the overlapping between them. We attempted to use a combination of visual attributes that leverage human perception system and do not impose cognitive and mental effort. The overall result of the evaluation shows that our work proved to be more effective than the current visualization.

## 1. INTRODUCTION

Spatial data and information about locations is most commonly represented on maps as this is most intuitive for viewers [1]. Day after day, maps and spatial data applications have become essential applications and data sources in different settings [2], such as political election analysis and weather forecasting. Although maps generally are not hard to read for even unsophisticated people, it is challenging to visualize the needed message in a way that anyone can grasp.

In this project, we worked with the University of Pittsburgh Graduate School of Policy and International Affairs Center for Metropolitan Studies (GSPIA). They are committed to developing the skills and strategies necessary to allow the primary policy holders of local governments to work across their boundaries to deal with the complex problems that arise at the local level. To this end, GSPIA has been at the vanguard of studying Regional Intergovernmental Organizations (RIGOs), political organizations constituted by the local governments within their respective regions that are supported by the active engagement of the region's community and citizens. These organizations differ significantly from how the federal government chooses to officially designate regional areas. Metropolitan Statistical Areas (MSAs), in contrast, are classified by the federal government based on commuting and commerce patterns. MSAs do not adhere to any local government. For more than two years, researchers at the Center for Metropolitan Studies have been reviewing websites, bylaws and other materials to create the most comprehensive database to date on RIGOs, in addition to population and demographic information. In GSPIA's continued efforts to explore regional governance, we are exploring a suitable approach to visualize overlap between and within the boundaries of RIGOs and MSAs. The main motivation for choosing this project is to gain experience working with an existing visualization, to enhance current features and to produce more concrete outcomes that will be beneficial for real users. Our core contribution is to find a way to visualize the complexity of a combined RIGOs and MSAs data by using visual encoding attributes that allow users to easily see the distinction in boundaries.

## 2. RELATED WORK

Visualizing spatial data is challenging due to the number of considerations, such as the shape, the position, and the borders of the location that the designer should address in order to maximize the amount of information that needs to be conveyed through the visualization. In our work, we have dealt with a number of different considerations: 1) Two different categorical datasets: RIGO and MSA, 2) Two versions of RIGOs, 3) The overlapping between RIGOs and MSAs 4) The population for each RIGO and MSA. Several papers attempted to tackle similar considerations by suggesting novel ideas based on the form of information that they want to be delivered.

In Malik et. al, [3] work, authors tried to show the temporal correlation in a choropleth map. The range of the temporal correlation that they showed was from -1.0 to 1.0. They divided the range, in terms of the visual encoding, into two different segments; above zero and below the zero. They used two distinct colors for each segment and a different saturation within each segment, to assist the reader to distinguish between the two different segments and the degree of the correlation within each segment. From this, we got inspired to have separate maps for RIGOs and MSAs beside having separate views: national and state view.

Furthermore, hierarchical data is commonly visualized in treemap layout [4,5,6], with the same color but different saturation. However, researchers [7] have used squarified treemap in order to use the grouping feature to visualize different hierarchical level. Thus, we have grouped each RIGOs and MSAs by line boldness while implying the census for each RIGOs and MSAs using different saturations will aid the users in distinguishing the boundaries between RIGOs and MSAs.

In a recent work [8], the author provides a framework to help the designers tackle situations involving comparison. The framework holds four components: 1) Identify the Comparative Elements. 2) Identify the Comparative Challenges. 3) Identify a Comparative Strategy. 4) Identify a Comparative Design. This framework provided support to our approach in solving the issue of having two different versions from RIGOs by using a different texture. In the next section, we will explain how we handled all these factors in details.

## 3. VISUALIZATION DESIGN

The collective motivation for our visualization design team and members of the GSPIA Center for Metro Studies was to create an easy to use and understand visualization that allows users to clearly distinguish difference between RIGOs and MSAs boundaries. More specifically, the visual will provide users with a representation of how counties choose to affiliate or separate themselves within these larger organizations. The aim of the visualization is to create a platform for researchers to draw insights about governance patterns in the US through an interactive visualization.

### 3.1 Data

We worked with two primary datasets. The first outlines cross-boundary data by organizations. Within this dataset, the fields (variables) in which we are most interested are the organizations, a RIGO code for each organization, the state to which the organization belongs, the population of that organization and the region type. The second dataset provides more insight into secondary RIGO affiliations with different organizations. Variables include the RIGO name and each of the RIGO codes.

### 3.2 Design Process

Identifying the best way to present data is subjective can be obscure to some extent [9]. Thus, knowing the audience for specific visualization can aid its creators to present it in a way that can be valuable. Since our main audience are faculty members and researchers from GSPIA, we considered the main users for this visualization to be sophisticated people with prior knowledge regarding: 1) How to use and read the maps. 2) The differences between counties and states. Next, we define the main issues that need to be solved: 1) How to differentiate between two categorical data: RIGO and MSA. 2) How to present the overlap between the two categorical data: RIGO and MSA. 3) How to present two versions from the same data category: RIGO_1 and RIGO_2. 4) How to implies the census for each: RIGO and MSA.

In our first meeting with GSPIA, we were introduced to the research team and their work on RIGOs and MSAs. They provided an overview of current visualization as well as the enhancements that they wanted us to implement. They proposed a set of visualizations which could be evaluated and implemented. We have evaluated their proposed solutions and found that some of them do not comply with the visualization principles that are found in the literature. For instance, one proposed solution included multiple colors that are used to encode the affiliations. That could impose mental efforts considering that the audience would need to remember these colors and their corresponding identifiers in the legend. The number of colors in the proposed visualization changes based on the number of RIGOs or MSAs affiliation in national or state view. Many colors do not convey the main message and may distract the audience. Since short-term memory is a limited resources and cannot consciously accommodate many visual attributes, it is important to consider such aspects while designing visualizations [10,11].

In creating a visualization that allows an audience to achieve their visualization goals, we must develop a way to encode the different RIGOs and their boundaries in a way that will be not be overwhelming to them.

### 3.3 Visual Encodings

In this project, our goal was to design effective visualizations based on visualization design considerations and principles, such as using preattentive visual attributes to encode data. Several research studies were conducted to identify preattentive visual attributes that leverage humans' visual processing system to make reading visualization easier [10]. The proper choice of the visual attributes will effectively support users' exploration tasks by making targets or patterns stand out. We considered and used several visual attributes that are classified as preattentive visual attributes, such as color, texture, and line boldness.

#### 3.3.1 Spatial Position

We encode the RIGOs and MSAs affiliations using choropleth map which is one type of the thematic maps ( figure 1 and figure 2 ). Each RIGO and MSA is mapped into its corresponding location on the map. Although other visualization techniques could be used to encode the same data, a choropleth map provide audience with affordance so that they can relate to the spatial position.

#### 3.3.2 Color and Saturation

Appropriate use of color while visualizing data is imperative since it significantly affects how visualization is being read and interpreted. While designing process, we identify two main categorical data: RIGOs and MSAs. We used two colors to map these categorical data and designed two distinct maps where each has its own distinct color ( figure 1 and figure 2 ). In addition to mapping the categorical data, we used color saturation to show the relational difference between the boundaries of the RIGOs and MSAs and to encode the population of RIGOs and MSAs. By using such visual attributes, users can spot those RIGOs or MSAs that have more population and perform their own analysis. Also, audience can identify the difference of adjacent RIOGs or MSAs if they have different saturations. One of the exploration tasks that users can achieve in our visualization is finding those RIGOs or MSAs that have more population and have small number of counties. In the view that compare RIGOs and MSAs affiliations, an additional third mixed color is used to encode both affiliations.

To promote the design consistency, color considerations are also used in the dashboard and font color. Brewer mentioned that color can be used for mapping categorical data, while the saturation can be used to map sequential or qualitative data [12]. Also, Iliinsky mentioned that color is appropriate to map categorical data[1].

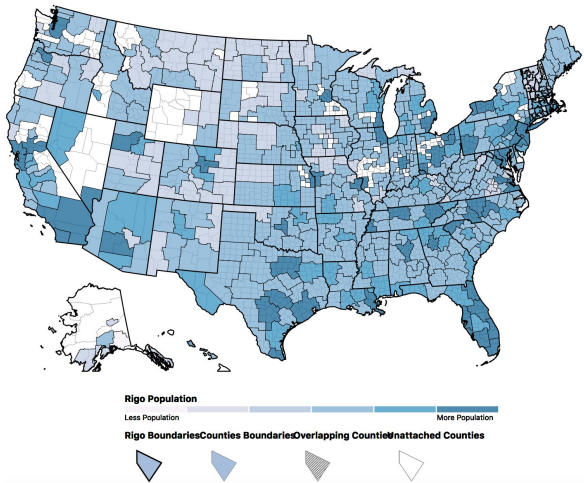

**Figure 1 : National map of RIGOs.** This view depicts RIGOs in the US where each RIGO is encoded with specific saturation that represents RIGO's population.

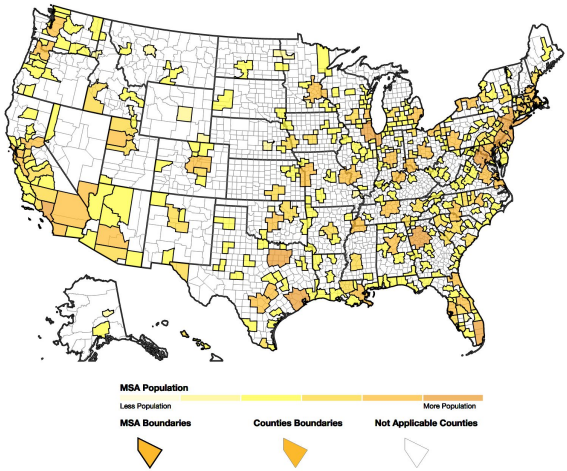

**Figure 2 : National map of MSAs.** This view depicts MSAs in the US where each MSA is encoded with specific saturation that represents RIGO's population. The color in this view is different from Figure 1 since it represent a distinct categorical data.

### 3.3.3 Line Boldness

In addition to the color visual attribute, we used line boldness to show the difference between the boundaries of states, RIGOs, and MSAs. For instance, the counties that have same RIGOs or MSAs affiliation are surrounded by a thicker line that show belonging relationship. Line boldness is visual attribute that could aid audience to differentiate between different groups that share same properties ( figure 1, figure 2, and figure 3 ).

### 3.3.4 Texture

We sought to approach the inherent visual complexity of a combined RIGO/MSA visual with a textural encoding solution that allows users to easily see the distinction in boundaries. We attempted to use a combination of two preattentive visual attributes to encode a secondary data which is about the counties that have two RIGOs affiliations. By using such combination, these counties will stand out, and users may perceive the distinction easily and quickly ( figure 3 ). Iliinsky recommend that patterns or textures can be used to map such nature of data [10,17].

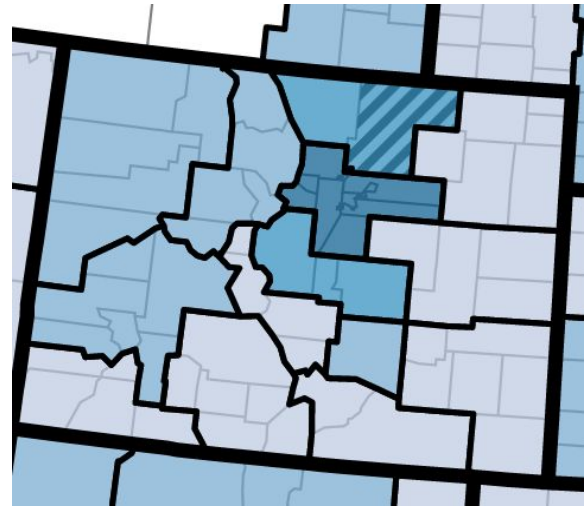

**Figure 3: Zoomed view of the national map that shows overlapping county that has two RIGOs affiliations.**

Each view of the map has a legend that clearly show what the visual encoding represents. Having a corresponding legend helps audience to be aware of how data is being encoded.

### 3.4 Interaction

A taxonomy has been offered by [13] for operations that must be considered in information visualization. In the previously described method of interaction, we are providing users with a way to manipulate the view and to navigate in order to examine patterns at a high level (national view) and detail at the lower level (state view). Additionally, revising the data fields to show more information for RIGOs or MSAs will be a matter of including them within the legend or tooltips upon hovering over the aforementioned RIGOs. Users may zoom, but we will also allow the ability to select a state to view its detailed boundaries. This ability to hover and select is also in line with view manipulation but more specifically, selecting items in order to highlight and filter them.

During the project, we worked with Javascript and D3 to mockup designs of the national and state views boundaries. We have designed a layout that provides a comprehensive view of the data at a national and state level. Throughout this process we created wireframes and mockups to present to our clients and decide on the best way to present these views to our users.

Early on in our project's development we focused on a few key areas to improve the visual. First, we proposed an idea for a dashboard component of the visual that would provide users with easy to understand figures that help illustrate different attributes related to RIGOs and MSAs like a comparative counts bar graph and a pie chart visualizing county affiliation trends. Although these data visualization approaches are far from novel, they were positively received by our client and have been validated as an effective approach to presenting a message with minimal confusion from users [14].

A second major improvement we worked towards was a reorganization of the presented data. We proposed moving from a zoomable United States map to a combination view that would allow users to view both the entire nation and a specific state on the same page in separate frames. In addition to this, we worked to develop a national map presentation that places less influence on county boundaries and provides greater visual presence to RIGO and MSA areas. Our GSPIA client viewed both of these efforts a good step towards improving the visualization.

After presenting our prototypes to the GSPIA partners for critique and validation, we were prepared to move forward with implementing our initial designs into an actual interactive visual.

## 3.5 Layout

We considered different layouts to apply to our visualization in order to provide the best view for our users. Our first implementation consisted of a view of a national map with the state map beneath it. We improved this by creating a side-by-side view of the national and state maps so that the user can easily compare boundaries (Figure 4). They have the ability to zoom in as well as the ability to select a RIGO view, MSA view, or both (as in Figure 5). In our final design, we have incorporated a dashboard that provides an overview of our data. In addition to a drop-down menu with instructions and explanations, we have the number of RIGOs and MSAs and an interactive chart to compare various affiliation categories.

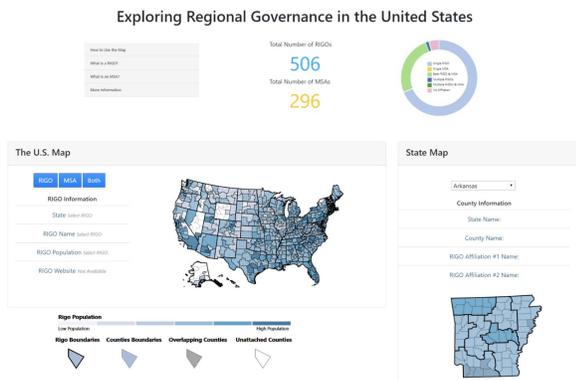

**Figure 4: Final Layout including a dashboard and side-by-side view of national and state maps.**

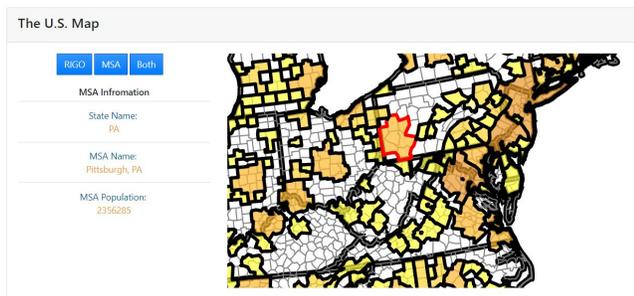

**Figure 5: Users are able to zoom-in and select view options for RIGO, MSA and Both**

## 4. TESTING IMPLEMENTATION

We wanted to test our design decisions and our approach at the visualization challenge that we have faced. The goal of our user study was to assess our users' ability to interact with our visualization and to interpret the information conveyed by the data we present. We also wanted to compare the interactions with the previous design and see if we have provided improvements.

### 4.1 Participants

Our 12 participants were undergraduate and graduate college students as well as working individuals who have completed degrees. Six participants completed the survey for the original design while the other six completed the survey for our design. This pool of participants is representative of our graduate student and young adult target users.

### 4.2 Design and Materials

The survey was a Google Form consisting of two main sections. The first section of our survey calls for the user to interact with the map and use it to answer multiple-choice questions based on the information offered in the designs. In answering these questions, participants provide a measure of accuracy and effectiveness for the design in question. For example, the main goal of our visualization is to allow viewers to compare the RIGOs and MSAs of the United States in terms of number and boundary. Therefore, our first question asks participants to identify whether the United States has more RIGOs or MSAs. Additionally, our design implementation aimed to further emphasize the overlap between RIGOs and MSAs. Our next questions asked, "Can a state have a RIGO that is also represented in another state?" and "Can a state have an MSA that is also represented in another state?". After answering these questions and becoming familiar with the design, the participants moved on to the next section.

The second section of this study takes participants through the System Usability Scale (SUS) created by John Brooke in 1986 [15]. The SUS is a ten-item Likert scale providing a numerical representation of subjective usability assessment.. The average SUS score is a 68 [16]. Therefore, in addition to providing a comparison to the original design, our results will yield a measure of the general usability of our design.

### 4.3 Results

In the preliminary section, every participant answered the first question correctly for each design. However, using the original design, participants were able to answer the second and third questions with only 33.33% and 66.67% accuracy respectively. Using our design, participants answered the second question with 50% accuracy and the third question with 66.67% accuracy.

To compare our design with the original design as a baseline, we performed a two-tailed two sample t-test of the user's SUS scores for each design. The results were not significant, however, the trend proved to be in favor of our design implementation. The usability scores of the original design ($M = 66.25$), $t(10)=-0.57$, $p>0.05$ were lower than the usability scores of our design ($M = 66.25$).

We then performed a one sample t test, comparing the SUS usability scores of our design with that of the average SUS score of 68.5. Statistically, the scores were not significantly different from the test value t($6$)=-0.22, $p>0.05$.

## 5. DISCUSSION

The preliminary results of our visualization showed that our design was more effective than the original design in allowing users to interpret the data. The participants who used our design answered these questions with more accuracy. Even though our SUS scores were not significantly different from the original design, there was a trend showing that scores were higher.

The study proved that there is room for improvement within the design. Our SUS scores, although not significant, showed a trend of being lower than the average score of 68. In regards to specific features, the participants pointed out their desire for a more responsive design, but were very appreciative of the instructions and comparison view that we have provided.

### 5.1 Advantages

The main advantage of this work is providing a suitable approach to visualize the complexity information due to the overlap between the boundaries of RIGOs and MSAs. The evaluation results show that the users have the ability to easily distinguishing the boundaries of RIGOs and MSAs. These significant advantages due to the features that we provided:

1. Separate maps for RIGOs, MSAs, and the overlap between RIGOs and MSAs called: both.

2. Two different views: State and national.

3. Zooming with hover feature in the national view.

4. Click feature that provide some information about the counties in the state view.

5. Filling the overlapping counties with texture as a visual encoding to make it distinguishable.

6. Different saturation based on the population for each RIGO and MSA.

7. A legend for each map to hand the user the key for the map.

8. An overall statistics to give the user a high-level picture about the data.

### 5.2 Limitations

As it is known that working in visualizing spatial data is usually surrounded with different limitations that might affect the final outcome, such as the size of the county. For example, some states have a huge number of small counties which make the differentiation process a little bit hard. Although working with a real client, GSPIA, was one of biggest motivation for us, we faced some difficulties to match their expectations and the best design practices.

### 5.3 Future Work[1]

Our future work will heavily focused on two main features:1) Improve the interactivity. 2) Develop a local search engine. 3) Having more responsive design that work with different platforms. 4) Improve the performance of the code.

---

[1] We already discussed with Jay Rickabaugh that we are going to continue working with him the first two months of the next semester.

## 6. MEMBER CONTRIBUTION

| Taks | Members |
| --- | --- |
| RIGOs map | Mohammed |
| MSAs map | Abdulelah |
| Both map | Mohammed and Abdulelah |
| State view | Mohammed and Abdulelah |
| Layout | Taylor |
| Evaluation | Taylor |
| Dashboard | Steve |
| Report | All |

## 7. ACKNOWLEDGMENTS

We would like to thank Jay Rickabaugh and the team at the University of Pittsburgh Graduate School of Public and International Affairs Center for Metropolitan Studies for all of their feedback and support.